\documentclass[aps,twocolumn,superscriptaddress,groupedaddress,nofootinbib]{revtex4}
\usepackage{graphicx}  
\usepackage{dcolumn}   
\usepackage{bm}        
\usepackage{amssymb,xcolor}   
\usepackage{hyperref} 
\usepackage{ulem}
\begin{document}

\title{Does Quarkonia Suppression serve as a probe for the deconfinement in small systems?}
\author{Partha Bagchi\footnote{parphy@niser.ac.in; parphy85@gmail.com}}

\address{School of Physical Sciences, National Institute of Science Education and Research, Jatni, Odisha -752050, India}

\author{Arpan Das \footnote{arpan.das@pilani.bits-pilani.ac.in}}

\address{Department of Physics, Birla Institute of Technology and Science, Pilani,  Rajasthan-333031, India}

\author{Ananta P. Mishra\footnote{apmishra@gmail.com}}

\address{School of Physical Sciences, National Institute of Science Education and Research, Jatni, Odisha -752050, India}
\date{\today}

\begin{abstract}

In high multiplicity proton-proton $(p-p)$ collisions, the formation of a deconfined state of quarks and gluons akin to Heavy Ion Collisions (HIC) has been a subject of significant interest. In proton-proton ($p-p$) collisions, the transverse size of the system is comparable to the longitudinal (Lorentz contracted) dimension, unlike the case in Nucleus-Nucleus ($A-A$) collision, leading to a hitherto unexplored effect of rapid decrease of temperature of the medium on quark-antiquark bound states. This allows us to probe a unique possibility of hadronization occurring before quarkonia dissociation within the medium. In small systems, a rapid change in temperature also introduces sudden changes in the Hamiltonian. This scenario prompts consideration of non-adiabatic evolution, challenging the traditional adiabatic framework. We demonstrate that non-adiabatic evolution may extend the longevity of quark-anti-quark bound states in $p-p$ collisions, even at higher multiplicities, offering new insights into the dynamics of strongly interacting matter produced in smaller collision systems.
\end{abstract}


\maketitle
\section{Introduction}
Dissociation of various quarkonia states is sensitive to the medium temperature which makes the Quarkonia suppression a probe for the presence of thermalized deconfined matter \cite{NA50:2000brc}. In the deconfined medium, the conventional mechanism for quarkonia suppression \cite{Matsui:1986dk} is the dissociation of quarkonia due to screening of the quark-antiquark potential in Quark-Gluon Plasma (QGP). Quarkonia can experience substantial yield modifications in the presence of QGP, primarily owing to Debye screening effects. When the temperature of the medium is higher than the dissociation temperature of the bound states, the potential between quark and anti-quark gets fully screened, and the states will no longer be bound states. Hence, the yield of those states will be suppressed. This picture implicitly assumes that quarkonia have enough time to respond to the medium, and this gives rise to the adiabatic evolution of quarkonia as a quantum state under the time-dependent Hamiltonian. The \textit{adiabatic} evolution refers to gradually changing conditions allowing the system to adapt its configuration. In such a process, a state corresponding to an initial Hamiltonian $H_0$ will evolve with time to the same eigenstate of the final Hamiltonian($H(t)$). 
The condition for adiabatic evolution is that the Hamiltonian undergoes gradual changes over time. This allows the initial eigenstates ample time to adjust in response to the evolving Hamiltonian, preventing any transitions to different eigenstates. Qualitatively, for adiabatic evolution the time scale corresponding to the change in Hamiltonian ($t_{\rm H}\sim\langle{m}|\dot{H}|n\rangle^{-1} $) is sufficiently higher than the time scale associated with the transition to the nearest eigenstate ($t_{\rm tr}\sim |E_m-E_n|^{-1}$), i.e., $t_{\rm H}>> t_{\rm tr}$~\cite{Bagchi:2015fry}. It is important to recognise that the change
in Hamiltonian stems from the dynamics of plasma evolution and is sensitive
to temporal variation of temperature. Depending upon the system under consideration the time evolution of the strongly interacting plasma can be quite rapid potentially leading to a situation where the condition $t_{\rm H} >> t_{\rm tr}$ may not always be satisfied. 

This necessarily demands a theoretical approach that incorporates a non-adiabatic evolution of bound states. Several attempts have been made to describe the evolution of quarkonia as non-adiabatic evolution (\cite{Dutta:2012nw, Bagchi:2014jya, Dutta:2019ntj, Bagchi:2018auv, Boyd:2019arx,Atreya:2014sea}). In these investigations, it has been argued that due to non-adiabatic evolution arising from the rapid temperature evolution, the initial quarkonium states can make a transition to different excited states and also to the continuum states in QGP. Interestingly the presence of a transient magnetic field, which is expected to be produced in noncentral heavy ion collision, can also give rise to a non-adiabatic quarkonia evolution(\cite{Bagchi:2023jjk, Bagchi:2018mdi, Iwasaki:2021nrz,Iwasaki:2021kms, Guo:2015nsa}). It is certain that the evolution must depend on the lifetime of QGP, mainly the temperature decay rate along with the medium's initial temperature. It is important to observe that for a rapid decrease in medium temperature, quarkonia may not get sufficient time to dissociate even if the initial temperature becomes more than the dissociation temperature.

Compelling data emerging from $p-Pb$ collisions at a center-of-mass energy of $\sqrt{s_{NN}} = 5.02$ TeV \cite{CMS:2014und} and from $p-p$ collisions at $\sqrt{s} = 5 \rm{TeV}$, $7 \rm{TeV}$, and $13 \rm{TeV}$ at the Large Hadron Collider (LHC) \cite{CMS:2016fnw}, reveals non-zero elliptic flow coefficients indicating the presence of a thermalized partonic medium. A recent study suggested that the suppression of quarkonia could be a signal of QGP formation in $p-p$ collisions \cite{Singh:2021evv}.  Considering the adiabatic evolution of quarkonia in a boost invariant system it has been argued in Ref.~\cite{Singh:2021evv} that if the quark-antiquark bound states are produced in a region where the effective temperature is higher than its dissociation temperature then the bound states will melt in the medium. Conversely, if the effective temperature is lower than the dissociation temperature, then the dissociation is minimal. We argue that in $p-p$ collisions small transverse size of the system can lead to a rapid decrease in temperature reducing the lifetime of the deconfined QCD medium.  Moreover in small systems, a rapid change in temperature can also introduce sudden changes in the Hamiltonian allowing for a non-adiabatic evolution. In this paper, we study the quarkonia suppression considering non-adiabatic evolution for small systems. We argue that non-adiabatic evolution and fast temperature decay can extend the longevity of quark-anti-quark bound states in $p-p$ collisions, even at higher multiplicities. 

The rest of the paper is organized in the following manner. In Sec.~\ref{formalism} we discuss the dissociation probability of quark-antiquark bound state under non-adiabatic evolution within the framework of time-dependent perturbation theory. This discussion is followed by the modeling of temperature evolution both in the pre-hydrodynamic stage as well as in the hydrodynamic stage. In Sec.~\ref{results} we present the main outcome of the paper, where we show that for small systems the dissociation of $J/\Psi$ can be suppressed primarily due to a shorter lifetime of the deconfined medium. In Sec.~\ref{summary} we conclude and summarize the results with an outlook to it. 

\section{Formalism}
\label{formalism}
\subsection{Time-dependent perturbation theory: non-adiabatic evolution}
Quarkonia are produced during the early stage (pre-equilibrium stage) of the collision. In a bottom-up thermalization approach, which is based on the QCD kinetic theory description, it can be argued that starting from an interacting out-of-equilibrium state one can achieve a thermalized medium at a later time $\tau_0$. In the absence of a thermalized medium, we can determine the initial state of quarkonia by solving the \textit{zero-temperature} Hamiltonian ($H_0 = \vec{p}^2/2M+\sigma r - \frac{4}{3}\alpha_s/r$)~\cite{Wong:1995jf}. Here $M$ denotes the reduced mass of the quark-antiquark system. However, as the medium achieves thermalization the zero-temperature Hamiltonian also evolves and transforms into its finite temperature counterpart denoted as $H= \vec{p}^2/2M+ \frac{\sigma}{\mu}(1 - \exp(-\mu r)) - \frac{4}{3}\alpha_s \exp(-\mu r)/r$~\cite{Karsch:1987pv}. Here $\alpha_s$ represents the strong coupling constant, $\sigma$ stands for the string tension, and $\mu$ represents the screening mass, which is temperature-dependent and determined by $\mu = \sqrt{6\pi\alpha_s} T$.  The time dependence in the Hamiltonian appears through the time dependence of temperature. As the system expands further the medium temperature eventually drops below the hadronization temperature, effectively reverting the Hamiltonian back to a zero-temperature state. We argue that the time evolution of the Hamiltonian can happen rapidly, within a time scale of the order of $1 ~\rm fm$ to $2 ~\rm fm$ time (denoted as $t_{\rm H}$) for small systems with a dominant transverse flow. However, the transition time scale $t_{\rm tr}$, from the energy difference of the quarkonia ground state to its next excited state, is around $1 ~\rm fm$ which is of the same order as $t_{\rm H}$ invalidating the adiabatic approximation.

Since this transition occurs quite rapidly, the initial quarkonia states evolve non-adiabatically and may undergo transitions to other states that are orthogonal to the initial ones. Hence, the probability that the original states make a transition to other orthogonal states can be calculated by generalizing the time-dependent perturbation theory method discussed by 
Lev Landau and Evgeny Lifshitz \cite{landau1991quantum,Bagchi:2015fry} for quantum systems. Let's begin by assuming that the initial state $|i\rangle$ is an eigenstate of the unperturbed Hamiltonian $H_0$ (at $\tau=0$), and it evolves to a generic state $|\psi\rangle$ in response to the perturbation $H'(\tau) = H(\tau) - H_0$. We aim to find the transition probability from the initial state to all other states orthogonal to a generic state $|m\rangle$ eigenstates of the unperturbed Hamiltonian. Here $|m\rangle$ could be a set of quantum states also. To achieve this, we first introduce a projection operator $Q = 1 - \sum_m |m\rangle\langle m|$, which projects out all states except for a few eigenstates represented by $|m\rangle$. Note that the initial states $|i\rangle$ can also belong to the set of $|m\rangle$ states. Therefore any state that is orthogonal to all states $|m\rangle$  can be expressed as $Q|\psi\rangle$. The transition amplitude $\mathcal{A}$ can be expressed as,
\begin{equation}
\label{eq:zeta1}
 \mathcal{A} = \langle\psi|Q|\psi\rangle.
\end{equation}
The evolved state $|\psi\rangle$ can be written in terms of the eigenstates $|n\rangle$ of the Hamiltonian $H_0$.
\begin{equation}
\label{eq:psi1}
|\psi\rangle = \sum_n c_n |n\rangle.
\end{equation}
The coefficients $c_n$ can be determined using perturbation theory, particularly from first-order perturbation theory~\cite{Bagchi:2015fry}:
\begin{eqnarray}
\label{eq:cnsudn}
 c_n \,&=&\, \delta_{ni}-i\Delta\tau\langle n|\left({1\over{\Delta\tau}}\int_0^{\Delta\tau} 
 H'(\tau)d\tau\right)|i\rangle  \nonumber \\
 \,&=&\, \delta_{ni}- i\Delta\tau\langle n|\bar{H'} |i\rangle .
\end{eqnarray}
Here, $\bar{H'}$ is defined as \footnote{In the integral above, within the first order of perturbations, we do not consider any leading order temporal dependence of states $|i\rangle$ and $|n\rangle$.}: 
\begin{equation}
\label{eq:vbar}
 \bar{H'} ={1\over{\Delta\tau}}\int_0^{\Delta\tau} H'(\tau)d\tau
\end{equation}
Using equations (\ref{eq:zeta1}), (\ref{eq:psi1}), (\ref{eq:cnsudn}), and (\ref{eq:vbar}), we arrive at the transition amplitude $\mathcal{A}$~\cite{Bagchi:2015fry}:
\begin{equation}
\label{eq:zeta2}
 \mathcal{A} = \Delta\tau^2 \langle\textbf{} i|\bar{H'}Q\bar{H'}|i\rangle
+\mathcal{O}((\Delta\tau\bar{H'})^3).
\end{equation}
It's important to note that the value of $c_n$ in Equation \ref{eq:cnsudn} is accurate up to the first order, which corresponds to the order of $\Delta\tau\bar{H'}$. Consequently, the value of $\mathcal{A}$ is accurate up to the second order in $c_n$, denoted as $(\Delta\tau\bar{H'})^2$.
Furthermore, we can express $\langle i|\bar{H'}Q\bar{H'}|i\rangle$ as:
\begin{eqnarray}
 \langle i|\bar{H'}Q\bar{H'}|i\rangle =  \langle i|{\bar{H'}}^2|i\rangle
 - \sum_m \langle i|\bar{H'}|m\rangle^2 .
\end{eqnarray}
This allows us to express $\mathcal{A}$ as:
\begin{equation}
\label{eq:zeta-transition}
\mathcal{A} = \Delta\tau^2 \left(\langle i|{\bar{H'}}^2|i\rangle - \sum_m \langle i|\bar{H'}|m\rangle^2\right).
\end{equation}
As stated above $\mathcal{A}$ quantifies the transition amplitude from the initial state to all other states that are orthogonal to all the states represented by $|m\rangle$. 
Certainly, for the present scenario, we consider the initial state $|i\rangle$ to be $|J/\Psi\rangle$ and the states in $|m\rangle$ to be $|J/\Psi\rangle$, $|\chi\rangle$, and $|\Psi'\rangle$, then the transition amplitude $\mathcal{A}$ represents the transition amplitude of $J/\Psi$ into 
 all the states other than $J/\Psi$, $\chi$, and $\Psi'$ in the Quark-Gluon Plasma (QGP). This transition amplitude for $J/\Psi$ can be expressed as,

\begin{eqnarray}
\label{eq:zeta-dis}
 \mathcal{A} &=& \Delta\tau^2 \bigg(\langle J/\Psi|{\bar{H'}}^2|J/\Psi\rangle - \langle J/\Psi|\bar{H'}|J/\Psi\rangle^2\\ \nonumber
 && ~~~~~~~~~- \langle J/\Psi|\bar{H'}|\chi\rangle^2- \langle J/\Psi|\bar{H'}|\Psi'\rangle^2\bigg). 
\end{eqnarray}
Using the above expression one can obtain $\Gamma\equiv|\mathcal{A}|^2$ which quantifies the dissociation probability of $J/\Psi$ in the QGP.

\subsection{Pre-equilibrium dynamics: evolution of effective temperature}
The key quantity that enters the expression of the transition probability is the perturbed Hamiltonian which carries the temporal dependence that originates from the time dependence of the system's temperature. modeling the time evolution of temperature is not trivial for the entire evolution of the plasma in heavy ion collision. Fortunately, hydrodynamic evolution plays a crucial role in the space-time evolution of the QCD medium after the partonic medium thermalizes starting from the pre-equilibrium stage. Irrespective of the initial stages, the hydrodynamic evolution unambiguously describes the bulk evolution of the medium. Therefore we can certainly choose a hydrodynamic evolution to model the temperature evolution or the evolution of the Hamiltonian. But to model the pre-equilibrium stages one may rely on the effective QCD kinetic theory description as has been discussed within the framework of bottom-up thermalization~\cite{Kurkela:2018oqw,Kurkela:2018xxd,Kurkela:2018vqr}. Qualitatively, in this approach, it has been argued that in non-expanding systems gauge bosons can rapidly achieve equilibrium (kinetic) among themselves followed by the equilibration of fermions. On the other hand, if the system undergoes a  rapid longitudinal expansion, partons may remain out of equilibrium, but the system can be effectively described by the fluid dynamics. Without going into the details of the model we consider the following ansatz for the proper time evolution of the \textit{effective} temperature ($T_{\rm{eff}}$) \footnote{It is effective temperature because the temperature is strictly defined in equilibrium only.}, 
\begin{equation}
\frac{T_{\rm{eff}}}{T_{\rm{Hydro}}}=\left(\frac{\tau}{\tau_{\rm{Hydro}}}\right)^{\frac{1}{7}\frac{\alpha-1}{(\alpha+3)}}
\label{equ9}
\end{equation}
The physical picture that prompts us to explore the above scaling is that the initial out-of-equilibrium partons scatter with each other to achieve the kinetic/thermal equilibrium. Therefore as the proper time approaches the thermalization time ($\tau_{\rm Them}$) system equilibrates as a whole. We also identify the thermalization time scale as the time scale when we can apply the hydrodynamic description ($\tau_{\rm{Hydro}}$). In principle,  all these different time scales can form a hierarchy, but we expect that if the thermalization is achieved very fast then the difference between different scales may not be too large not affecting the system dynamics significantly. The parameter $\alpha$ enters the above equation because the effective temperature can be defined through the $\alpha$-th moment of the Fermionic or Bosonic distribution function~\cite{Kurkela:2018oqw}. Physically the parameter $\alpha$ determines how fast the system achieves hydronization (onset of hydrodynamic description) or thermalization.  In the subsequent discussion, we appropriately choose $T_{\rm{Hydro}}$, $\tau_{\rm{Hydro}}$, and $\alpha$ to model the pre-equilibrium dynamics. Instead of going into the microscopic description, naively one can also assume the temperature starts at zero at some initial time and increases linearly until it reaches a value $T_{\rm{Hydro}}$ at time $\tau_{\rm{Hydro}}$. Such simple approximations can be useful to calculate the average perturbation (as described in Equation \ref{eq:vbar}) in the pre-thermalization stage.

\begin{figure}
\begin{center}
\includegraphics[width=0.5\textwidth]{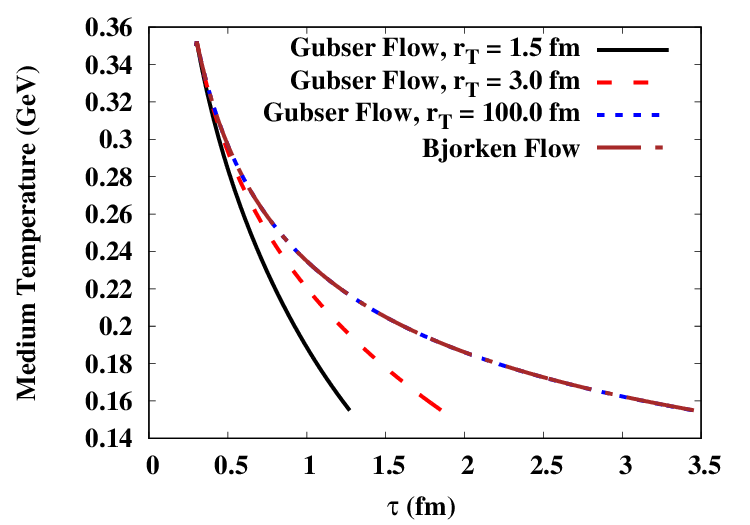}
\caption{Proper time ($\tau$) evolution of temperature ($T$) for different values of the transverse size ($r_T$) of the medium. All lines represent the temperature evolution following the Gubser flow without any viscous corrections. For comparison, we also show the temperature evolution as obtained from the Bjorken flow in the absence of viscosity. It is clear that with a large system size ($r_T=100$ fm) Gubser flow solution boils down to the Bjorken flow solution. With a smaller system size the lifetime of the deconfined medium decreases. In hydrodynamic evolution, we consider the temperature at the center of the transverse plane.}
\label{fig:tau_t_0}
\end{center}
\end{figure}

\begin{figure}
\begin{center}
\includegraphics[width=0.5\textwidth]{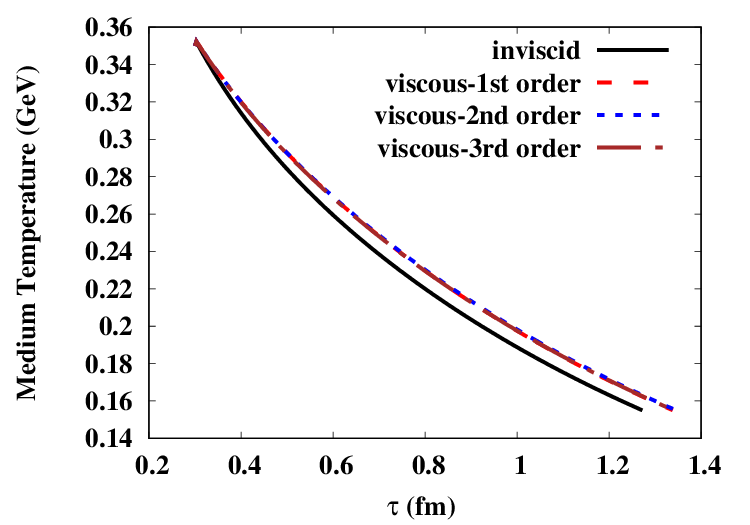}
\caption{Effect of viscous corrections on the evolution of the medium temperature. In this case, we consider $4\pi\eta/s=1$, $r_T=1.5$ fm. In this plot, we have considered viscous corrections up to third-order corrections. It is clear that viscous correction slows down the decrease of temperature with proper time increasing the lifetime of the deconfined medium. Here first order means Naiver-Stokes limit where $\pi$ is not an independent hydrodynamic variable. For the second order case $\hat{\chi}=0$ and for the third order $\hat{\chi}\neq 0$. }
\label{fig:tau_t_order}
\end{center}
\end{figure}

\subsection{Gubser flow and the associated temperature evolution}
Once we have a description of the pre-equilibrium effective temperature that also quantifies the pre-equilibrium dynamics of the Hamiltonian, now we can look into the temperature evolution due to the hydrodynamic flow dynamics. Notably, in $p-p$ collisions, the size of the produced medium is expected to be quite small, approximately $1.5$ fm \cite{McLerran:2013oju} and transverse expansion can not be ignored, which is otherwise neglected in the Bjorken flow solution for large systems.  To take into consideration the transverse expansion of the system in the present calculation we look into the  Gubser flow, first explored by Gubser and Yarom \cite{Gubser:2010ui}. This approach combines a ``boost-invariant" longitudinal flow, akin to the Bjorken flow, with consideration for transverse flow. The evolution of key thermodynamic quantities, including energy density ($\epsilon$) and shear stress ($\pi$), within the framework of Gubser flow with third-order viscous corrections, is detailed in \cite{Chattopadhyay:2018apf,Dash:2020zqx}.

\begin{eqnarray}
 \frac{d\hat\epsilon}{d\rho} &=& -\left(\frac{8}{3}\hat\epsilon-\hat{\pi} \right)\tanh(\rho) \label{eq:e_evolution}
\\   \frac{d\hat\pi}{d\rho}  &=& -\frac{\hat\pi}{\hat\tau_{\pi}}+\tanh(\rho)\left(\frac{4}{3}\hat\beta_{\pi}-\hat\lambda\hat\pi- \hat\chi\frac{\hat{\pi}^2}{\hat\beta_{\pi}}\right)\label{eq:pi_evolution}
\end{eqnarray}
 The dimensionless quantities, $\hat\epsilon$ and $\hat\pi$, are expressed as   $\hat\epsilon= \hat{T}^4 =\epsilon \tau^4 = 3\hat{P}$ and $\hat\pi=\pi\tau^4$ where $\tau$ is the proper time and $\hat{T}$ is related to temperature.
The parameters has been chosen \cite{Chattopadhyay:2018apf} as $\epsilon = \frac{3}{\pi^2} T^4$, $\hat\tau_\pi (=c/\hat{T}) $ is related to relaxation time, where $c=5\frac{\eta}{s}$, $\hat\beta_\pi= 4\hat{P}/5$, $\hat{\lambda}=46/21$ and the third order correction parameter $\hat\chi=72/245$. 
 The conformal time $\rho$ can be written as 
 \begin{equation}
    \rho = -\sinh^{-1}\left(\frac{1-q^2\tau^2+q^2x_T^2}{2q\tau}\right)
 \end{equation}
 where  $q$ is an arbitrary energy scale, which is related to the transverse size of the medium ($r_T$) like $q=\frac{1}{r_T}$, $x_T$ is the position in the transverse plane. One can retrieve the Bjorken flow solution by taking the limit $r_T\rightarrow \infty$ or $q\rightarrow 0$. One can also use the $(3+1)$ dimensional hydrodynamic description for a more accurate description of non-boost invariant flow with nontrivial rapidity dependence. But considering the possible boost invariance in ultra-relativistic collisions we restrict ourselves to the analytically solvable hydrodynamic description with transverse expansion.

To demonstrate the effect of the transverse expansion on the evolution of temperature we solve the  
evolution equations (\ref{eq:e_evolution}) and (\ref{eq:pi_evolution}) with initial conditions $ T = T_{\rm{Hydro}} = 350$ MeV and $\hat\pi = \frac{4}{3}\hat\beta_{\pi}\hat\tau_{\pi}$ at $\tau = \tau_{\rm{Hydro}} = 0.3$ fm for various system sizes ($r_T$). The results as shown in Fig.(\ref{fig:tau_t_0}) indicate that, as $r_T$ increases, the lifetime of the Quark-Gluon Plasma (QGP) increases. This is because the time scale over which the temperature ($T$) falls just below $T_c$ (the QCD phase transition temperature which is considered as $\sim$ 150 MeV) increases with the increase in transverse size. At sufficiently large $r_T$, the variation of temperature ($T$) with proper time ($\tau$) for Gubser flow closely resembles that of Bjorken flow. 
Additionally, from Fig.~\ref{fig:tau_t_order} we may observe that the decay of temperature is slower for the viscous case compared to the inviscid case (Fig. \ref{fig:tau_t_0}) for identical initial conditions. In Fig. \ref{fig:tau_t_order}, it is evident that the ($T$ vs. $\tau$) plot for different orders of viscous corrections largely overlaps with each other for the lower bound of $\frac{\eta}{s}$ (equal to $\frac{1}{4\pi}$). In contrast, the inviscid case consistently exhibits faster temperature decay compared to all viscous scenarios. The variation of temperature with system size clearly indicate that for small system the time evolution of the system can be rapid as compared to the large systems allowing us to explore the scenario of non-adiabatic evolution.

\begin{figure}[]
\begin{center}
\includegraphics[width=0.5\textwidth]{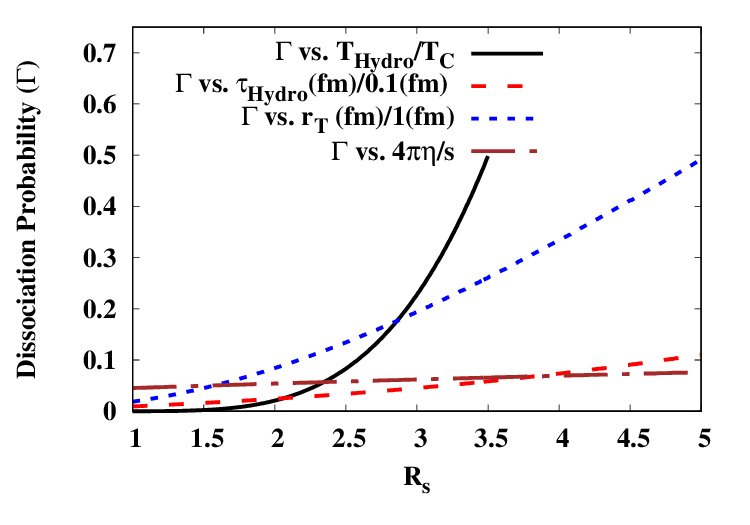}
\caption{Dissociation Probability of ${J/\Psi}$ as a function of the dimensionless quantity $R_s$ (see text for a detailed description). $R_s$ indicates $T_{\rm Hydro}/T_c$ ratio for the black line, $\tau_{\rm{Hydro}}(\rm fm)/0.1(\rm fm)$ for the red line, $r_T(\rm fm)/1(\rm fm)$ for the blue line, and $4\pi\eta/s$ for the brown line. In this case, the evolution of effective temperature in the pre-equilibrium stage has been modeled using the power law in Eq.(\ref{equ9}) with $\alpha=2$. We do not find any significant change in the results for different values of $\alpha$, e.g., $\alpha=3$ and 4. We have taken the suitable value of $
\alpha$ as given in Ref.~\cite{Kurkela:2018oqw}.}
\label{fig:prob_viscous_alph}
\end{center}
\end{figure}

\begin{figure}
\begin{center}
\includegraphics[width=0.5\textwidth]{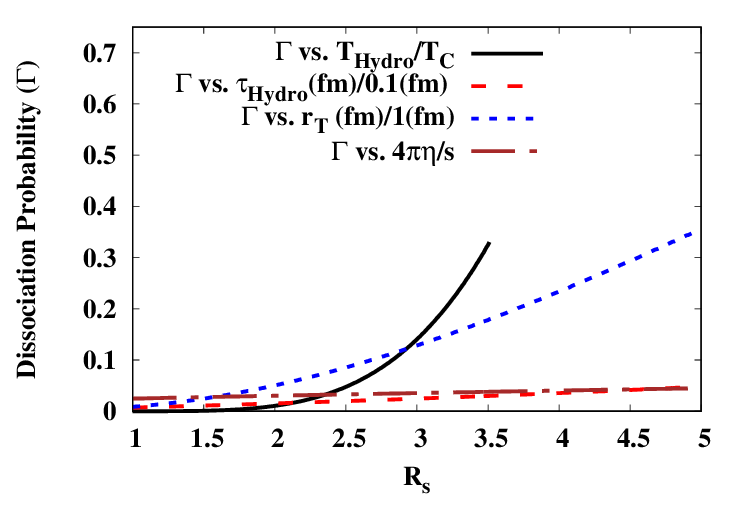}
\caption{Dissociation Probability of ${J/\Psi}$ as a function of the dimensionless quantity $R_s$ (see text for a detailed description). $R_s$ indicates $T_{\rm Hydro}/T_c$ ratio for the black line, $\tau_{\rm{Hydro}}(\rm fm)/0.1(\rm fm)$ for the red line, $r_T(\rm fm)/1(\rm fm)$ for the blue line, and $4\pi\eta/s$ for the brown line. In this case, the evolution of effective temperature in the pre-equilibrium stage has been modeled using the linear rise of effective temperature starting from zero to $T_{\rm{Hydro}}$.}
\label{fig:prob_viscous}
\end{center}
\end{figure}

\section{results and discussions}
\label{results}
The factors that can affect the dissociation probability of $J/\Psi$ are (1) the evolution profile of effective temperature in the pre-hydronization/ pre-thermalization stage, (2) the initial thermalization/hydronization temperature ($T_{\rm{Hydro}}$), (3) the time scale for hydronization ($\tau_{\rm Hydro}$), (4) the transverse size of the system, and (5) viscous correction to hydrodynamic flow. As mentioned above in this work we have considered two different scenarios for the evolution of the temperature in the pre-hydronization stage. One is the power law profile as indicated in Eq.(\ref{equ9}) and the other one is a linear rise of temperature from zero to $T_{\rm{Hydro}}$. The results for the power law profile and for the linear profile have been shown in Fig.(\ref{fig:prob_viscous_alph}), and    Fig.(\ref{fig:prob_viscous}) respectively. In both figures to see the effect of $T_{\rm{Hydro}}$ on the dissociation probability, we vary $T_{\rm{Hydro}}$ from $T_c$ to $3T_c$ for a fixed $\tau_{\rm{Hydro}}=0.3$ fm, $4\pi\eta/s=1$, and $r_T=1.5$ fm. On the other hand, to observe the effect of  $\tau_{\rm{Hydro}}$ on the dissociation probability in both figures we consider $T_{\rm{Hydro}}=350$ MeV,  $4\pi\eta/s=1$, $r_T=1.5$ fm, and vary $\tau_{\rm{Hydro}}$ within the range of $0.1$ fm to $0.5$ fm. Similarly, we keep $T_{\rm{Hydro}}=350$ MeV, $\tau_{\rm{Hydro}}=0.3$ fm  $4\pi\eta/s=1$ fixed and vary $r_T$ within the range $1.0-5.0$ fm to demonstrate the effect of the transverse size of the system. Finally, to isolate the effect of viscosity on the $J/\Psi$ dissociation probability in both figures we only vary the shear viscous coefficient $4\pi\eta/s$ within the range $1-5$, keeping all the other parameters fixed, i.e., $T_{\rm{Hydro}}=350$ MeV, $\tau_{\rm{Hydro}}=0.3$ fm, $r_T=1.5$ fm.  We choose this set of parameters to determine the highest possible dissociation of $J/\Psi$. Moreover in our calculations, we have assumed $J/\Psi$ is produced at the centre of the medium ($x_T=0$), with no initial transverse momentum $(p_T)$ facing the QGP medium up to hadronisation. By setting up this scenario, we aim to maximize the time the produced $J/\Psi$ spends within the QGP, which in turn should maximize the probability of it undergoing dissociation. 

In Figs.\ref{fig:prob_viscous_alph} and \ref{fig:prob_viscous} we plotted the dissociation probability with respect to the dimensionless ratio $R_s$, where $R_s$ indicates $T_{\rm Hydro}/T_c$ ratio for the black line, $\tau_{\rm{Hydro}}(\rm fm)/0.1(\rm fm)$ for the red line, $r_T(\rm fm)/1(\rm fm)$ for the blue line, and $4\pi\eta/s$ for the brown line.
Looking at these figures, we can observe some important trends. First, the dissociation probability tends to increase gradually with both $\tau_{\rm{Hydro}}$ and viscosity. However, this increase is not very steep. The dependence of the dissociation probability on $\tau_{\rm{Hydro}}$ is rather convoluted. Because both the pre-equilibrium and the hydrodynamic stages depend on $\tau_{\rm{Hydro}}$. 
On the other hand, the dissociation probability rises quite rapidly as the system size ($r_T$) increases. This system size scaling of the $J/\Psi$ dissociation probability indicates that for small systems quark-antiquark bound states can survive more with respect to the large systems (if we ignore any dissociation mechanism due to hadronic scattering in hot and dense hadron gas). This is predominantly because for larger $r_T$ the lifetime of the deconfined medium is larger. Furthermore, the rate of increase becomes even more pronounced with higher viscosity ($\eta/s$). This is because the viscous effect tends to reduce the rate of fall of temperature due to hydrodynamic expansion. However, it's important that even with these increases, as long as the transverse size of the system remains less than $3$ fm, the dissociation probability stays of the order of $20\%$. For a system size of order 1.5 fm the dissociation probability can be as small as $10\%$. Therefore even in the presence of deconfined matter the dissociation probability of $J/\Psi$ is not significantly large for small systems. From these plots, one can also conclude that the dissociation probability can be sensitive to the modeling of the effective temperature for the pre-equilibrium dynamics.

We emphasize that we have calculated the probability of dissociation of $J/\Psi$ with no transverse momentum and produced at the center of the medium ($x_T=0$). However, the $J/\Psi$ can have finite $p_T$ and it can be produced anywhere in the transverse plane. In this scenario, the produced $J/\Psi$ will face QGP for a much shorter time and can get out of the medium well before hadronization. Consequently, the probability of $J/\Psi$ dissociation in such cases will be significantly lower compared to our previous calculations. Therefore, the overall probability of dissociation may be insignificant in these more realistic conditions.

\section{Summary}
\label{summary}
In this paper, we estimate an upper bound on the $J/\Psi$ dissociation probability for small systems. With realistic modeling of the pre-hydronization stage along with a hydrodynamic evolution with the transverse expansion we showed that the survival probability of quark-antiquark bound states can be significantly large in small systems even for large multiplicity. Due to a finite transverse system size in small systems, non-adiabatic evolution can affect the dissociation probability of $J/\Psi$. Therefore quarkonia suppression may not be a clear signature for deconfinement in small systems even with large multiplicity. The technique we've employed to calculate the dissociation probability of $J/\Psi$ has broader applications beyond high-energy physics. It can be adapted for use in other fields like atomic or molecular physics and chemistry. For example, one can easily generalize this approach to calculate the ionization of hydrogen-like atoms caused by electric impulses. Moreover, this technique can be readily extended to calculate the ionization of atoms or molecules when exposed to an electromagnetic pulse, making it a versatile tool in various scientific domains. 

\section*{Acknowledgements}
We thank the organizers of the ICPAQGP 2023 at Puri, India, and the 2nd workshop on the `Dynamics of QCD matter' organized at the National Institute of Science Education and Research Bhubaneswar (NISER), India for the kind hospitality and for creating the opportunity for fruitful discussions and further development related to this work. In these conferences, this work has been presented by APM and PB. We would also like to acknowledge Ashutosh Dash, Amaresh Jaiswal, Surasree Mazumder, Sukanya Mitra, Tamal K. Mukherjee, and Victor Roy for illuminating discussions.

\bibliographystyle{apsrev4-1} 
\bibliography{qqbib}{}

\end{document}